\documentclass[10pt,showpacs,showkeys,aps, amsmath, amssymb, aps, pre,twocolumn,groupedaddress,superscriptaddress,nofootinbib,longbibliography,author-year]{revtex4-1}

\usepackage[colorlinks,linkcolor=black,citecolor=blue,urlcolor=black]{hyperref}
\usepackage[english]{babel}
\usepackage{balance}
\usepackage{natbib}

\usepackage{amsmath}
\usepackage{amsfonts,amssymb}
\usepackage{enumerate,pifont}
\usepackage{graphicx}
\usepackage[update,prepend]{epstopdf}
\usepackage{mathptmx}      

\newtheorem{theorem}{Theorem}

\newcommand{\al}{\alpha}

\newcommand{\eps}{\varepsilon}
\newcommand{\ph}{\varphi}

\newcommand{\trn}{^{\rm\scriptscriptstyle T}}

\newcommand{\mC}{{\mathbb C}}
\newcommand{\mR}{\mathbb R}

\newcommand{\inr}{\!\in \mR}

\newfont{\bfb}{msbm10 scaled 1000}

\newcommand{\bmat}[1]{\left [ \begin{array}{#1}}
\newcommand{\emat}{\end{array}\right ] }

\renewcommand{\geq}{\geqslant}
\renewcommand{\le}{\leqslant}
\renewcommand{\ge}{\geqslant}

\DeclareMathOperator{\sign}{sign}

 \DeclareMathOperator{\sat}{sat}

\DeclareMathOperator{\diag}{diag}

\def\CheckPDFoutput{%
\CheckPDFoutput%
\ifx\unprotect\undefined%
        \DeclareGraphicsRule{.jpg}{bmp}{}{}%
\else%
        \pdfoutput=1%
        \fi%
}%

\begin{document}

\title{Hidden attractors in aircraft control systems with saturated inputs}


\author{B.R.~Andrievsky}
\email{boris.andrievsky@gmail.com}%
\affiliation{
Institute of Problems in Mechanical Engineering, the Russian Academy of Sciences, Saint Petersburg, Russia}
\affiliation{Mathematics and Mechanics Faculty, Saint Petersburg State University, Russia}
\affiliation{ITMO University, St. Petersburg, Russia}
\author{E.V.~Kudryashova}
\email{e.kudryashova@spbu.ru}
\affiliation{Mathematics and Mechanics Faculty, Saint Petersburg State University, Russia}
\author{N.V.~Kuznetsov}
  \email{nkuznetsov239@gmail.com}
\affiliation{Mathematics and Mechanics Faculty, Saint Petersburg State University, Russia}
\affiliation{Department of Mathematical Information Technology,
University of Jyv\"{a}skyl\"{a}, Jyv\"{a}skyl\"{a}, Finland}
\author{O.A.~Kuznetsova}
  \email{olga.kuznetsova@spbu.ru}%
\affiliation{Mathematics and Mechanics Faculty, Saint Petersburg State University, Russia}
\author{G.A.~Leonov}
  \email{g.leonov@spbu.ru}%
\affiliation{Mathematics and Mechanics Faculty, Saint Petersburg State University, Russia}
\affiliation{
Institute of Problems in Mechanical Engineering, the Russian Academy of Sciences, Saint Petersburg,  Russia}

\begin{abstract}
In the paper, the control problem with limitations on the
magnitude and rate of the control action in aircraft control
systems, is studied. Existence of hidden limit cycle oscillations in the
case of actuator position and rate limitations  is demonstrated by the examples of piloted aircraft pilot involved oscillations (PIO) phenomenon and the airfoil flutter suppression system.
\end{abstract}


\maketitle

\section{Introduction}

In the paper, the control problem with limitations on the magnitude and rate of the control action in  aircraft control systems, is studied.
This problem has long attracted the attention of scientists and developers of flight control systems, is still a challenging one, without losing its relevance at the present time.

It is shown in the literature that in the motion of an aerodynamically stable aircraft
a stable limit cycle with a small amplitude and an unstable one with a large amplitude can coexist. If the aircraft is aerodynamically unstable, then one of the two stable limit cycles with a small amplitude can be realized. In addition, there is also an unstable limit cycle, the presence of which makes it necessary to study the stability of an aircraft with an automatic control system ``in large'', i.e. when large perturbations (including those from the side of a pilot) are applied to the aircraft, they can lead him beyond the amplitude of the unstable limit cycle.

An influence of non-linearities like a ``saturation'' in a pilot--aircraft loop is commonly treated as a possible origin of the so-called {\it Pilot Involved  Oscillation} (PIO), which leads to serious degrade of the piloted aircraft performance, up to the stability loss and the aircraft destruction, cf.  \cite{Dornheim92,Shifrin93}. This phenomenon is characterized by rapidly developing oscillations with increasing amplitude at angular velocities, overloads and angular movements of the manned vehicle. The main non-linear factor leading to this phenomenon is, as a rule, the limitation of the speed of deviation of the aircraft control elements (aerodynamic control surfaces of the aircraft), which can lead to a delay in the response of the aircraft to the pilot's commands. The study of transient regimes with such a motion leads to the need to develop a mathematical theory of global analysis of control systems of aircraft.

As the results obtained,  for studying the processes that can arise in nonlinear control systems of aircraft (including nonlinear oscillations), simple computer modeling is an unreliable tool that can lead to wrong conclusions.

The remainder part of the paper is organized as follows. The analytical-numerical method for hidden oscillations localization is briefly recalled in Sec.~\ref{Sec:analmeth}. The adopted  dynamic actuator model with saturations in the magnitude and rate is given in Sec.~\ref{Sec:modact}. In Sec.~\ref{Sec:modact} hidden oscillations in the pilot-aircraft loop (related to the PIO phenomenon) are studied and localized by means of the iterative analytical-numerical method. Hidden oscillations in the unstable aircraft angle-of-attack control system with saturated actuator are demonstrated in Sec.~\ref{sec:angle}. Section \ref{sec:PIO} is devoted to hidden oscillations in the pilot-aircraft loop. Hidden limit cycle oscillations in the airfoil flutter feedback suppression system are
studied in Sec.~\ref{Sec:flut}. Concluding remarks and the future work intentions are given in
Sec. \ref{Sec:conc}. Some background information is given in Appendix \ref{appaer}.

\section{Analytical and numerical analysis of oscillations in nonlinear control systems}\label{Sec:analmeth}

Further we consider a nonlinear control system with one scalar non-linearity in the Lur'e form
 \begin{equation}\label{sys_gen}
   \dot {\bf x} = {\bf P}{\bf x}+{\bf q}\psi({\bf r}\trn {\bf x}),\quad
   {\bf x}\in \mathbb{R}^n,
 \end{equation}
where ${\bf P}$ is a constant $(n\times n)$-matrix,
${\bf q}, {\bf r}$ are constant $n$-dimensional vectors,
$\trn $ denotes transpose operation,
$\psi(\sigma)$ is a continuous piecewise-differentiable
scalar function, $\psi(0)=0$.

One of the main tasks of the investigation of nonlinear dynamical models \eqref{sys_gen}
is the study of established (limiting) behavior of the system after the transient processes are over, i.e., the problem of localization and analysis of \emph{attractors} (limited sets of system's states, which are reached by the system from close initial data after transient processes).
For numerical localization of an attractor one needs to choose an initial point in the basin of attraction and observe how the trajectory, starting from this initial point, after a transient process visualizes the attractor.
Thus, from a computational point of view, it is natural
to consider the following classification of attractors being
either \emph{self-excited} either \emph{hidden}
based on the simplicity of finding the basins of attraction in the phase space \cite{LeonovK-2013-IJBC,KuznetsovL-2014-IFACWC,LeonovKM-2015-EPJST,Kuznetsov-2016}:
\emph{self-excited attractor} has the basins of attraction touching
an unstable stationary point, thus,
can be revealed numerically by the integration of trajectories\footnote{
Remark that in numerical computation of trajectory over a finite-time interval
it may be difficult to distinguish a \emph{sustained oscillation} from a \emph{transient oscillation}
(a transient oscillating set in the phase space,
which can nevertheless persist for a long time).
},
started in small neighborhoods of the unstable equilibrium,
while \emph{hidden attractor} has the basin of attraction,
which does not touch equilibria, and is hidden somewhere in the phase space
\cite{LeonovK-2013-IJBC,KuznetsovL-2014-IFACWC,LeonovKM-2015-EPJST,Kuznetsov-2016}.
For example, hidden attractors in systems \eqref{sys_gen}
correspond to the case of \emph{multistability} when the stationary point is
stable and coexist with a stable periodic orbit, or there are several coexisting periodic orbits.
The \emph{classification of attractors as being hidden or self-excited}
was introduced by in connection with the discovery of the first hidden Chua attractor
\cite{KuznetsovLV-2010-IFAC,LeonovKV-2011-PLA,BraginVKL-2011,LeonovKV-2012-PhysD,KuznetsovKLV-2013,KiselevaKKKLYY-2017,StankevichKLC-2017}
and has captured much attention
(see, e.g. \cite{BurkinK-2014-HA,LiSprott-2014-HA,LiZY-2014-HA,PhamRFF-2014-HA,
ChenLYBXW-2015-HA,KuznetsovKMS-2015-HA,SahaSRC-2015-HA,SemenovKASVA-2015,SharmaSPKL-2015-EPJST,ZhusubaliyevMCM-2015-HA,WeiYZY-2015-HA,
DancaKC-2016,JafariPGMK-2016-HA,MenacerLC-2016-HA,OjoniyiA-2016-HA,PhamVJVK-2016-HA,RochaM-2016-HA,WeiPKW-2016-HA,Zelinka-2016-HA,
BorahR-2017-HA,BrzeskiWKKP-2017,FengP-2017-HA,JiangLWZ-2016-HA,KuznetsovLYY-2017-CNSNS,MaWJZH-2017,MessiasR-2017-HA,SinghR-2017-HA,VolosPZMV-2017-HA,WeiMSAZ-2017-HA,ZhangWWM-2017-HA}).

\subsection{Analysis of oscillations
by the harmonic balance and describing function method}

The describing function method (DFM) is a widely used engineering method
for the search of oscillations which are close to the harmonic periodic oscillations.
This method is {\it not strictly mathematically justified}
and is one of the approximate methods of analysis of oscillations
(see, e.g. \cite{KrylovB-1947,PopovPaltov63,Khalil-2002,LiuDowellThomas_JFS05}).

Let us recall a classical way of applying the DFM.
Introduce a transfer function
\(
	W(s) = {\bf r}\trn  \big({\bf P} - s {\bf I} \big)^{-1} {\bf q}
\),
where $s$ is a complex variable, ${\bf I}$ is a unit matrix.
In order to find a periodic oscillation,
a certain coefficient of harmonic linearization $k$
is introduced in such a way that the matrix
${\bf P}_0 = {\bf P} + k{\bf qr}\trn $
has a pair of pure-imaginary eigenvalues $\pm i\omega_0 (\omega_0>0)$.
The numbers $\omega_0 > 0$ and $k$ are defined by
\begin{equation}\label{eq:omega0k}
	\omega_0: \ {\rm Im}\, W( i \, \omega_0) = 0, \quad
	k = - \big({\rm Re}\, W(i \, \omega_0) \big)^{-1}.
\end{equation}
If such $\omega_0$ and $k$ exist,
then system \eqref{sys_gen}
has a periodic solution ${\bf x}(t)$ for which
\(
	\sigma(t) = {\bf r}\trn {\bf x}(t) \approx a_0 \cos \omega_0 t.
\)
Consider $\varphi(\sigma)=\psi(\sigma)-k\sigma$,
then, following the DFM,
the amplitude $a_0$ of oscillation can be obtained from the equation
\begin{equation}\label{df_crit}
      \Phi(a) = \int_0^{2\pi/\omega_0} \ph\big(a\cos(\omega_0 t)\big) \cos(\omega_0 t) dt, \quad
      \Phi(a_0)=0,
\end{equation}
where $\Phi(a)$ is called a \emph{describing function}.
Assume that $r = (1 ,r_2..,r_n)$, then
one can use  the following initial data
\begin{equation}\label{dfmic}
  {\bf x}(0) = (a_0,0,...,0)
\end{equation}
for numerical localization of an attractor.
However it is known that classical DMF
can suggest the existence of non-existing periodic oscillation
(see, e.g. \cite{Shahgildyan-1972,KudryashovaKLYY-2017})
and initial data \eqref{dfmic} does not necessarily lead
to the localization of an attractor.

To get initial data in the basin of attraction of an attractor,
we can use the following modification of the DFM \cite{LeonovK-2013-IJBC}.
Let us change $\varphi(\sigma)$ by $\varepsilon\varphi(\sigma)$, where
$\varepsilon$ is a small parameter,
and consider the existence of a periodic solution for system
\begin{equation} \label{sys_gen_phi_0}
   \dot {\bf x} ={\bf P_0}{\bf x}+\varepsilon{\bf q}\varphi({\bf r}\trn {\bf x}).
\end{equation}
Consider a linear non-singular transformation\footnote{Such transformation exists for non-degenerate transfer functions} ${\bf x}={\bf S}{\bf y}$, such that
system \eqref{sys_gen_phi_0} is transformed to the form
\begin{equation} \label{sys_hlin}
 \begin{aligned}
 & \dot y_1\!=\!-\omega_0y_2\!+\!\varepsilon b_1\varphi(y_1\!+\!{\bf c}_3\trn {\bf y}_3),~
 \dot y_2\!=\!\omega_0y_1\!+\!\varepsilon b_2\varphi(y_1\!+\!{\bf c}_3\trn {\bf y}_3),
\\
 & \dot {\bf y}_3 = {\bf A}_3{\bf y}_3+\varepsilon {\bf b}_3\varphi(y_1+{\bf c}_3^{*} {\bf y}_3),
 \end{aligned}
\end{equation}
where $y_1$, $y_2$ are scalars, ${\bf y}_3$, ${\bf b}_3$ and ${\bf c}_3$ are $(n-2)$-dimensional vectors,
$b_1$ and $b_2$ are scalars;
${\bf A}_3$ is a constant $((n-2)\times(n-2))$ matrix
all eigenvalues of which have negative real parts.
\begin{theorem}\label{th_stable}(\cite{LeonovK-2013-IJBC})
If there exists a number $a_0>0$ such that
\begin{equation}\label{det_a0}
  \Phi(a_0)=0, \quad b_1 \Phi'(a_0) < 0,
\end{equation}
then system \eqref{sys_gen_phi_0} has a stable periodic solution with initial data
\begin{equation}\label{dfm-ic}
	{\bf x}(0) = {\bf S}\,\big(a_0 + O(\varepsilon), \, 0, \, {\bf O}_{n-2}(\varepsilon)\big)\trn .
\end{equation}
\end{theorem}

Remark that there are known examples where DFM cannot reveal the existing oscillation
(see, e.g. \cite{Tzipkin-1984,BraginVKL-2011,LeonovK-2013-IJBC}).
For example,  well-known Aizerman's and Kalman's conjectures
on the absolute stability of nonlinear control systems
are valid from the standpoint of DFM,
while there are known various counterexamples of nonlinear systems
where the only equilibrium, which is stable,
coexists with a hidden periodic oscillation
(see, e.g. \cite{Pliss-1958,Fitts-1966,Barabanov-1988,BernatL-1996,LeonovBK-2010-DAN,BraginVKL-2011,LeonovK-2011-DAN,LeonovK-2013-IJBC};
the corresponding discrete examples are considered in \cite{Alli-Oke-2012-cu,HeathCS-2015}).
In this case the DFM can be justified and analog of Theorem 1
can be obtained for a special class of nonlinearities only \cite{LeonovK-2013-IJBC}.
For example \cite{Fitts-1966,LeonovK-2013-IJBC,LeonovM-2017-DAN},
system with transfer function
\begin{align*}
W(s)=\frac{s^2}{\big((s+0.03)^2+0.09^2\big)\big((s+0.03)^2+1.1^2\big)}
\end{align*}
has infinite sector of linear stability,
but, e.g., for nonlinearity $\psi(\sigma) = \sign(\sigma)$
a hidden attractor can be found numerically.

\subsection{Localization of hidden oscillations by the continuation method and describing function methods}

Consider an analytical-numerical procedure for the hidden attractors localization based on the continuation method and DFM.
For that we construct a finite sequence of functions
\[
  \varphi^{j}(\sigma) = \varepsilon^{j}\varphi(\sigma),
  \quad \varepsilon^{j}=j/m \mbox{ \ or\ } \varepsilon^{j}=m/j, j=1,..,m
\]
such that system \eqref{sys_gen} with initial function $\varphi^1(\sigma)$
has a nontrivial attractor $\mathcal{A}^1$,
which either is self-excited and can be visualized by the initial data from small vicinity of one of the equilibria
either can be visualized by the initial data \eqref{dfm-ic}.
On each next step of the procedure (i.e. for $j\geq 2$),
the initial point for a trajectory to be integrated
is chosen as the last point of the trajectory integrated on the previous step.
Following this procedure and sequentially increasing $j$,
two alternatives are possible:
the points of $\mathcal{A}^j$ are in the basin of attraction
of attractor $\mathcal{A}^{j+1}$,
or while passing from system \eqref{sys_gen} with the function $\varphi^j(\sigma)$
to system \eqref{sys_gen} with the function $\varphi^{j+1}(\sigma)$,
a loss of stability bifurcation is observed and attractor $\mathcal{A}^j$ vanishes.
If, while changing $j$ from $1$ to $m$,
there is no loss of stability bifurcation of the considered attractors,
then an attractor for the original system with $\varphi(\sigma)$
(at the end of the procedure) is localized.

\section{Modeling dynamical actuator with saturations}\label{Sec:modact}

In the  linear settings, an aircraft actuator is usually modeled by the second-order differential equation as
\begin{align}
\ddot\delta(t)+2\xi_\text{act}\omega_\text{act}\dot\delta(t)+\omega^2_\text{act}\delta(t)=\omega^2_\text{act}u(t),
\label{bia31}
\end{align}
or, in the Laplace transform representation, as
\begin{align}
W_\text{act}(s)=\left\{\dfrac{\delta}{u}\right\}=\dfrac{\omega^2_\text{act}}{s^2+2\xi_\text{act}\omega_\text{act} s+\omega_\text{act}^2},
\label{bia32a}
\end{align}
where $u$ denotes the commanded controlling surface (elevator) deflection, $\delta$ stands for the actual deflection, $\omega_\text{act}$ $\xi_\text{act}$ are the actuator natural frequency and damping ratio, respectively, $s\in\mC$ denotes the Laplace transform variable.

\begin{figure}[htpb!]
\centering
\includegraphics[width=1\linewidth]{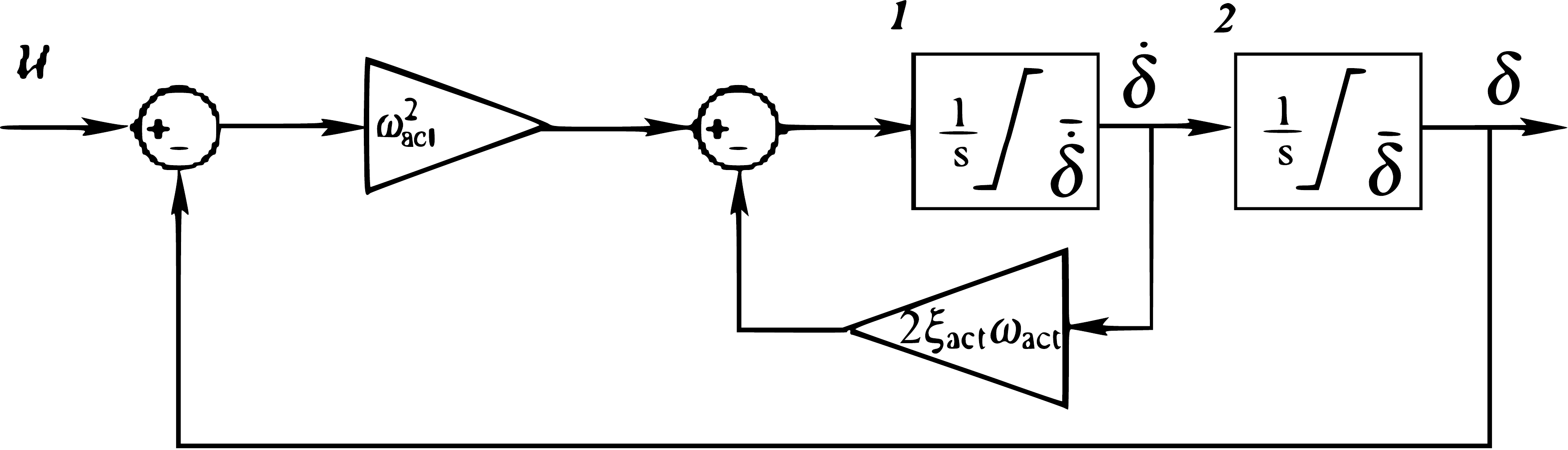}
\caption{Block diagram of nonlinear actuator model with magnitude and rate limitations, cf. \cite{BiannicTarbouriech_06,BiannicTarbouriech09}.}
\label{bianact}
\end{figure}

In the present study, more realistic actuator model involving both
magnitude and rate limitations is considered instead of the linear one \eqref{bia31}. Following \cite{BiannicTarbouriech_06,BiannicTarbouriech09}, let us replace \eqref{bia31} by the nonlinear model shown in  Fig.~\ref{bianact}. This nonlinear model involves two limited integrators
which are dynamic nonlinearities. Integrator {\bf 1}  is associated to the rate limitation
 $\bar{\dot{\delta}} $, while Integrator {\bf 2} corresponds to the magnitude limitation  $\bar{{\delta}}$.

The limited nonlinear dynamic integrator with input $x(t)$, output $y(t)$ and the saturation level $\bar y$ (i.e. $|y|\le \bar y$ for all $t$; we assume that the limitations are symmetrical) is described by the following model:
\begin{align}
\dot y=\begin{cases}0,&~~\text{if}~~(y\ge \bar y)\cap(x\cdot y>0), \\
x,&~~\text{otherwise},
\end{cases}
\label{bia32c}
\end{align}
where $\cap$ denotes the logical ``AND'' operation.

In the Lur'e form, nonlinear integrator \eqref{bia32c} may be approximately described by the following model \cite{BiannicTarbouriech_06,BiannicTarbouriech09}:
\begin{align}
\dot\sigma=x-\lambda(\sigma-y),\quad y=\sat_{\bar y}\sigma,
\label{bia32b}
\end{align}
where gain $\lambda>0$ is sufficiently large.  More precisely, as follows from [Lemma 5.1]\cite{BiannicTarbouriech09} for any continuous input
signal   $x(t)$  with bounded derivatives, the approximation error tends to zero when the tuning parameter $\lambda$ increases.

Based on the saturated integrator description given by \eqref{bia32b}, the actuator model pictured in Fig.~\ref{bianact} is represented by the following equations:
\begin{align}
\begin{cases}
\dot\sigma_1=\omega^2_\text{act}(u-\delta)-
2\xi_\text{act}\nu-\lambda(\sigma_1-\nu),\cr
\dot\sigma_2=\nu-\lambda(\sigma_2-\delta),\cr
\nu=\sat_{\bar{\dot \delta}}\sigma_1.
\end{cases}
\label{eqND16}
\end{align}

\section{Unstable aircraft angle-of-attack control system with saturated actuator}\label{sec:angle}
Consider the following linearized model of short-term dynamics of an unstable aircraft
in the vertical plane as in \cite{BiannicTarbouriech09}:
\begin{align}
\begin{cases}
\dot\al(t)=-Y_\al\al(t)+q(t)-Y_{\delta}\delta_e(t),\\
\dot q(t)=M_\al\al(t)+M_qq(t)+M_\delta\delta_e(t),
\end{cases}
\label{nd23}
\end{align}
where $al$, $q$ and $\delta_e$ denote the angle-of-attack, the pitch-rate and the elevator deflection,
respectively; $Y_\al$, $Y_\delta$, $M_\al$, $M_q$, $M_\delta$ are the aircraft model parameters which
are assumed to be constant during the considered time slot.
Let the following proportional feedback law be realized by means of the aircraft
longitudinal stability augmentor:
\begin{align}
u(t)=K_\al\al(t)-K_\text{cw}X_\text{cw}(t),
\label{nd24}
\end{align}
where $K_\al$, $K_\text{cw}$ are stability augmentor gains, $X_\text{cw}$ is the control wheel deflection,
produced by a pilot.
In the sequel, model \eqref{nd23} parameters are taken as follows (see \cite{BiannicTarbouriech09}): $Y_\al = 0.47$,
$Y_\delta =0.16$, $M_\al =0.82$, $Mq =-0.43$, $M_\delta =-4.4$ (in SI units). Note that $M_\al >0$ and,
therefore, the considered aircraft model is weathercock unstable with respect to the
angle of attack. Parameter $K_\al$ of stability augmentor \eqref{nd24} is taken as $K_\al = 0.35$.

In the present example, the actuator dynamics are described by \eqref{eqND16} taking into
account limiting the maximum speed of moving the aircraft controlling surface. Such
a limitation may appear due to the saturation of the of hydraulic fluid supply to the
cylinder when the supplying channels are fully opened. The actuator position limitation
is assumed to be sufficiently large, therefore it is neglected. To simplify the
exposition, without loss of generality, gain $K_\text{cw}$ in \eqref{nd24} is taken equal to one. The actuator
model parameters (in SI units) are taken as $\omega_\text{act} = 20$, $\xi_\text{act} = 0.6$, $\dot{\bar \delta} = 10/57.3$.

\subsection{Localization of hidden oscillations}
\begin{figure*}[htpb!]
\centering
\includegraphics[width=0.9\linewidth]{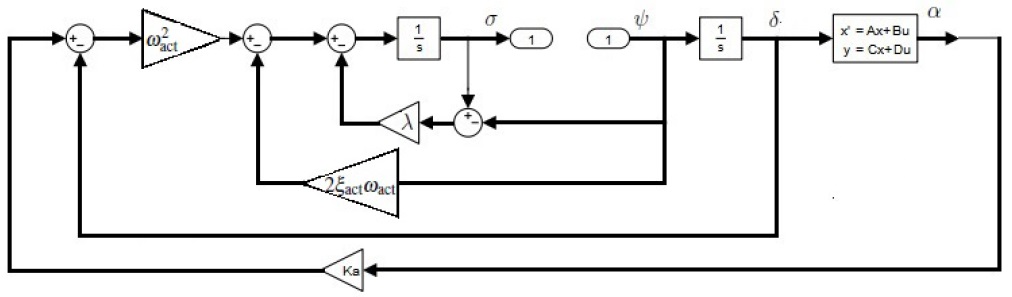}
\caption{Simulink block diagram of the linear part of the system \eqref{eqND16}, \eqref{nd23}.}
\label{figND5}
\end{figure*}

\begin{figure}[htpb!]
\centering
\includegraphics[width=1\linewidth]{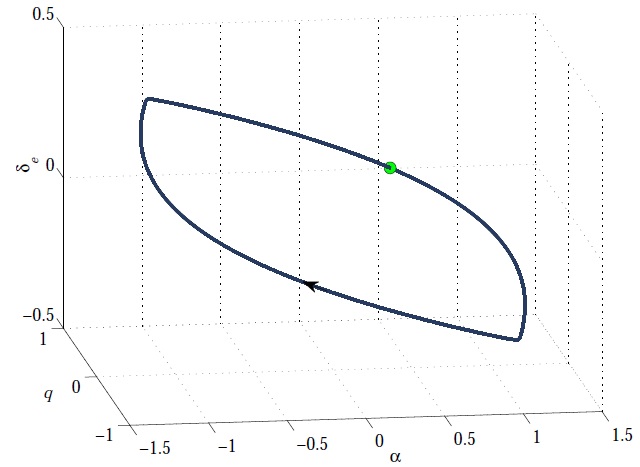}
\caption{Projection of the limit cycle in system \eqref{eqND16}, \eqref{nd23},
 \eqref{nd24} to the subspace
 $(\al, q, \delta)$. Initial point ${\bf x}(0)=\big(-0.1745201, 0.009257612$, $0.5652268, 0.9543608\big)\trn $
(marked by a circle).}
\label{figND6}
\end{figure}

\begin{figure}[htpb!]
\centering
\includegraphics[width=1\linewidth]{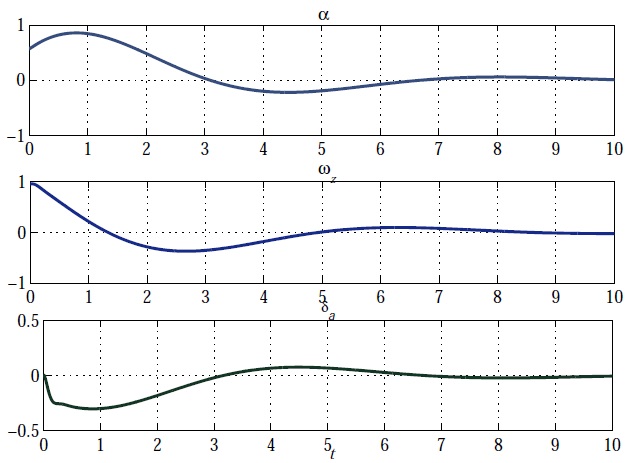}
\caption{Time histories of  $\al$, $q$, $\delta$ in the system \eqref{bia31}, \eqref{nd23} with linear actuator model.}
\label{figND7}
\end{figure}

\begin{figure}[htpb!]
\centering
\includegraphics[width=1\linewidth]{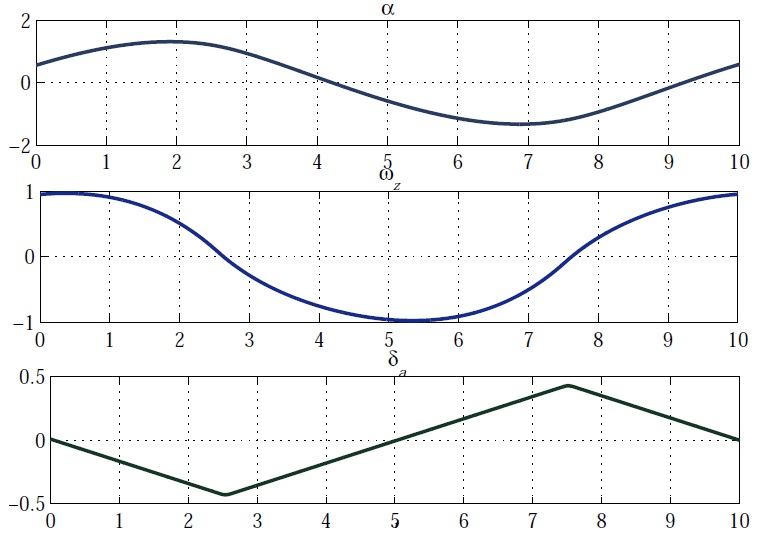}
\caption{Time histories  of  $\al$, $q$, $\delta$ in the system with nonlinear actuator model.}
\label{figND8}
\end{figure}

Let us apply to the considered system the method of hidden oscillations localization,
described in Sec. \ref{Sec:analmeth}, cf. \cite{Bragin_TiSU11,Dudkowski_PR16,LeonovKuznetsov_IJBC13,LeonovK-2011-IFAC,LeonovKV-2011-PLA,LeonovKV-2012-PhysD}. To this end represent the system model \eqref{eqND16}, \eqref{nd23},
 \eqref{nd24} (assuming $u_\text{p}(t)\equiv 0$) to the state-space form \eqref{sys_gen}, considering $\psi(\sigma)$ as an
input of the linear part of the system, and $\sigma$ as its output. Introduce the state-space
vector ${\bf x}$ as ${\bf x} = (\sigma, \delta_e, \al,q)\trn \inr^4$, $n=4$. This leads to the following matrices in \eqref{sys_gen}:
\begin{align}
&{\bf P}\! =\!\begin{pmatrix}\!-\lambda&\!-\!\omega^2_\text{act}&K_\al\omega^2_\text{act}&0\\
0& 0& 0& 0\\
0&-Y_\delta&-Y_\al& 1\\
0&M_\delta&M_\al&M_q
\end{pmatrix},~~
{\bf q}\!=\!\begin{pmatrix}\lambda\!-\!2\xi_\text{act
}\omega_\text{act}\\
1\\0\\0
\end{pmatrix},\notag\\
&{\bf r}=\begin{pmatrix}1&0&0&0
\end{pmatrix}\trn .
\label{nd25}
\end{align}
The block diagram of the linear part of the system is demonstrated in Fig.~\ref{figND5} {\bf(}on this diagram, the Simulink {\sl State-Space block} represents model~\eqref{nd23}{\bf)}.
For the given above parameter values ($Y_\al = 0.47$, $Y\delta = 0.16$, $M_\al = 0.82$, $M_q =
-0.43$, $M_\delta = -4.4$, $K_\al = 0.35$, $\omega_\text{act} = 20$, $\xi_\text{act} = 0.6$, $\lambda = 100$) one gets
\begin{align}
&{\bf P}\! =\!\begin{pmatrix}\!-100&\!-\!400&-140&0\\
0& 0& 0& 0\\
0&-0.16&-0.47& 1\\
0&-4.4&0.82&-0.43
\end{pmatrix},~~
{\bf q}\!=\!\begin{pmatrix}76\\
1\\0\\0
\end{pmatrix},\notag\\
&{\bf r}=\begin{pmatrix}1&0&0&0
\end{pmatrix}\trn .
\label{nd26}
\end{align}
Application the iterative procedure of Sec.~\ref{Sec:analmeth} shows existence in the system the
limit cycle oscillation with the initial point ${\bf x}(0) =\begin{pmatrix}\sigma(0)&
\delta_e(0)&\al(0)&q(0)\end{pmatrix}\trn =
\begin{pmatrix}-0.1745201& 0.009257612& 0.5652268& 0.9543608\end{pmatrix}\trn $. Projection of the limit cycle
to the subspace $(\al, q, \delta)$ is depicted in Fig.~\ref{figND6}. Time histories of variables $\al$, $q$, $\delta$
for given ${\bf x}(0)$ are plotted in Figs.~\ref{figND7}, \ref{figND8}. Figure \ref{figND7} represents the case when no rate
limitations are taken into account {\bf(}i.e. the case of a linear actuator model \eqref{bia31}{\bf)}, while
Fig. \ref{figND8} refers to the case of nonlinear actuator model \eqref{eqND16} and reflects an influence of
the rate limitations to the overall system behavior.
\section{Aircraft-pilot loop}\label{sec:PIO}
The problem of control with limitations on the magnitude, rate, and energy of the control action due to its relevance for practice has attracted the attention of scientists and developers of automatic control systems for aircrafts for a long time, see \cite{AndrievskyKL-2017,BriegerKerr_ACC07,BriegerKerr_ACC08,AviationSafety97, QueinnecTarbouriech_IC06, TarbouriechTurner09} and the references therein.

Influence of non-linearities like the ``saturation'' can also cause the so-called {\it Pilot Involved Oscillations} (PIOs), which violate the aircraft piloting, cf. \cite{Powers84,AviationSafety97,PachterMiller_CS98,BriegerKerr_ACC07,BriegerKerr_ACC08,BriegerKerrPostlethwaite_JCD12, BriegerKerr_JGCD12,AndrievskyKLP-2013-IFAC, AcostaYildiz14,AndrievskyKL-2015,AndrievskyKL-2017}. This phenomenon is characterized by rapidly developing fluctuations with increasing amplitude of angular velocities, accelerations, and angular movements of manned aircraft. Despite the nature of PIO is not completely clear, it is generally recognized that the main factor leading to PIO is limitations of the rate of deviation
of the aircraft control inputs (such as the controlling aerodynamic surfaces). This restriction may result in a delay in the response of the aircraft to the pilot commands.

As noted in \cite{AviationSafety97,Duda_ASTE98}, PIOs usually arise in situations where the pilot is trying to maneuver by aircraft with a high precision. The study of transient regimes with this motion leads to the need of development a mathematical theory of global analysis of flight control systems.

Below the hidden PIO-like oscillations in aircraft-pilot contour are studied based by the example of piloted research aircraft {\it X-15}.
 The transfer function of {\it X-15} research aircraft longitudinal dynamics from the elevator deflection $\delta_e$ to pitch angle $\theta$ is taken as follows \cite{AlcalGordillo_ACC04, AmatoIervolino_ECC01,AmatoIervolino_CDC00,AndrievskyKL-2017, MehraPrasanth_CCA98}:
\begin{align*}
G_\delta^\theta(s)\!=\!\left\{\dfrac{\theta}{\delta_e}\right\}\!=\!
\dfrac{3.48(s+0.883)(s+0.0292)}{(s\!+\!0.3516)(s\!+\!0.02845)\big(s^2\!+\!1.68s\!+\!5.29\big)},
\end{align*}
where $\delta_e(t)$ denotes the elevator deflection with respect to the trimmed value, $\theta(t)$ stands for the pitch angle (all variables are given in the SI units), $s\in\mC$.

The pilot is usually modeled as a serial element in the closed-loop system. Having enough flight skills, the pilot develops a stable relationship between his control action and a specific set of flight sensor signals \cite{McRuer67}. Based on \cite{BarbuReginatto_ACC99, LoneCooke_AEST14,McRuer67} the following pilot model in the form of a lead-lag-delay unit is taken in the present study:
\begin{align}
G_\text{p}(s)=\left\{\dfrac{u}{\Delta\theta}\right\}=
K_\text{p}\dfrac{T_Ls+1}{T_Is+1}\text{e}^{-\tau_es},
\label{eqND18}
\end{align}
where $\Delta\theta$ is the displayed error between desired pitch angle $\theta\trn $ and actual one $\theta$; $u(t)$ denotes the pilot's control action, applied to the elevator servo; $K_\text{p}$ is the pilot
static gain; $T_L$ is the lead time constant; $T_I$ stands for the lag time constant; $\tau_e$ denotes the effective time delay, including
transport delays and high frequency neuromuscular lags. Since the present paper is focused to studying autonomous systems behavior, it is assumed in the sequel that $\theta\trn \equiv 0$ and, therefore,  $\Delta\theta\!=\!- \theta$.

To apply the method of \cite{LeonovKV-2012-PhysD,Bragin_TiSU11,LeonovKuznetsovYuldahsev_DM11,LeonovKuznetsov_IJBC13}, the time-delay transfer function $\exp{-\tau_es}$ in pilot model \eqref{eqND18} is approximated employing the first-order {\it Pade} $(1,1)$ {\it representation} \cite{GolubLoan89}
as  $\text{e}^{-\tau_es}\approx\dfrac{-\tau s+2}{\tau s+2}$. This leads to the following second-order model of the pilot dynamics:
\begin{align}
G_\text{p}(s)=\left\{\dfrac{u}{\Delta\theta}\right\}=
K_\text{p}\dfrac{(T_Ls+1)(-\tau s+2)}{(T_Is+1)(\tau s+2)}.
\label{eqND19}
\end{align}

Finally, the transfer function of the open-loop aircraft-pilot system from elevator deflection $\delta_e$ to the pilot's control action $u$ has the following form:
\begin{align}
G_\text{p}(s)=\left\{\dfrac{u}{\delta_e}\right\}=
K_\text{p}\dfrac{(T_Ls+1)(-\tau s+2)}{(T_Is+1)(\tau s+2)}\,G_\delta^\theta(s).
\label{eqND20}
\end{align}
In the present study the pilot model parameters are taken as: $K_\text{p} = 1.8$, $T_L = 0.6$~s, $T_I = 0.2$~s, $\tau = 0.2$~s. Consequently, one obtains the following transfer function $G(s)$:
\begin{align}
&G_\text{p}(s)=-\dfrac{10.428(s-10)(s+1.667)}{(s+10)(s+5)(s+0.3516)}\notag\\
&\quad\times\dfrac{(s+0.883)(s+0.0292)}{(s+0.02845)\big(s^2+1.68s+5.29\big)}.
\label{eqND21}
\end{align}

\begin{figure}[htpb!]
\centering
\includegraphics[width=1\linewidth]{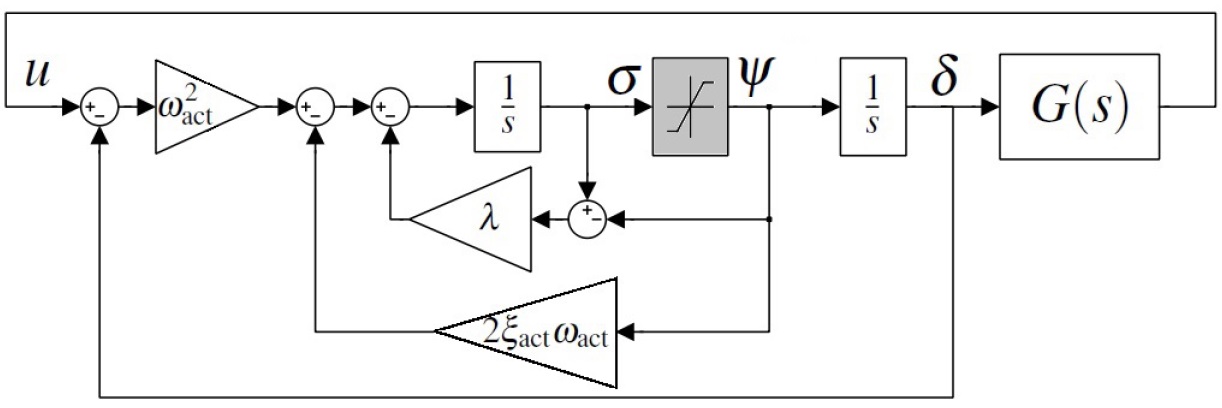}
\caption{Block-diagram of the closed-loop aircraft-pilot system  \eqref{eqND16}, \eqref{eqND21} with rate saturation.}
\label{figND2}
\end{figure}

The actuator is modeled as a second-order dynamical unit with a rate limitation \eqref{eqND16}. The actuator model parameters are taken as $\omega_\text{act} = 50$~rad/s, $\xi_\text{act} = 0.6$, $\bar{\dot\delta}_\text{act}=
15/57.3$~rad/s, $\lambda= 100$~s$^{-1}$. Block-diagram of the closed-loop aircraft-pilot system \eqref{eqND16}, \eqref{eqND21} with saturated actuator model is pictured in Fig.~\ref{figND2}.

\subsection{Localization of hidden oscillations}
Let us apply the procedure of hidden oscillations localization to piloted aircraft control system \eqref{eqND16}, \eqref{eqND21}. To this end consider the linear subsystem with input $\psi$ and output $\sigma$ (see. Fig.~\ref{figND2}) and, following Sec.~\ref{Sec:analmeth}, represent it in the state-space form \eqref{sys_gen}.
In the considered case, $n=8$ and in a certain basis of state space variables,  for given above numerical values of the model parameters, one obtains the following matrices in \eqref{sys_gen}:
\begingroup\makeatletter\def\f@size{7}\check@mathfonts
\begin{align*}
& {\bf P}\!=\!\begin{bmatrix}\vspace{-1mm}
-100\hspace{-2.6mm}& 0\hspace{-2.6mm}& 4.69\cdot 10^4\hspace{-2.6mm}& -\!3.48\cdot  10^5\hspace{-2.6mm}& -\!1.14 \cdot 10^6 \hspace{-2.6mm}&-\!7.24\cdot 10^5 \hspace{-2.6mm}&-\!2.02 \cdot 10^4\hspace{-2.6mm}& -\!2500\\\vspace{-1mm}
0&-\!17.1 &-\!86.8 &-\!194 &-\!327 &-\!102 &-\!2.65& 1\\\vspace{-1mm}
0&1&0 &0 &0&0 &0& 0\\\vspace{-1mm}
0&0&1 &0 &0&0 &0& 0\\\vspace{-1mm}
0&0&0 &1 &0&0 &0& 0\\\vspace{-1mm}
0&0&0 &0 &1&0 &0& 0\\\vspace{-1mm}
0&0&0 &0 &0&1 &0& 0\\\vspace{-1mm}
0&0&0 &0 &0&0 &0& 0\vspace{2mm}
\end{bmatrix}\!,
\end{align*}
\endgroup
\begin{align}
&{\bf q}\!=\!\big[40,\! 0,\! 0,\! 0,\! 0,\! 0,\! 0,\!-\!1\big]\trn,~~ {\bf r}\!=\!\big[1, 0, 0, 0, 0, 0, 0, 0\big]\trn.
\label{eqND22}
\end{align}

If the equilibria of system \eqref{sys_gen} are stable,
for the numerical search of hidden oscillations
we can use the continuation method of Sec.~\ref{Sec:analmeth}.

\begin{figure}[htpb!]
\centering
\includegraphics[width=1\linewidth]{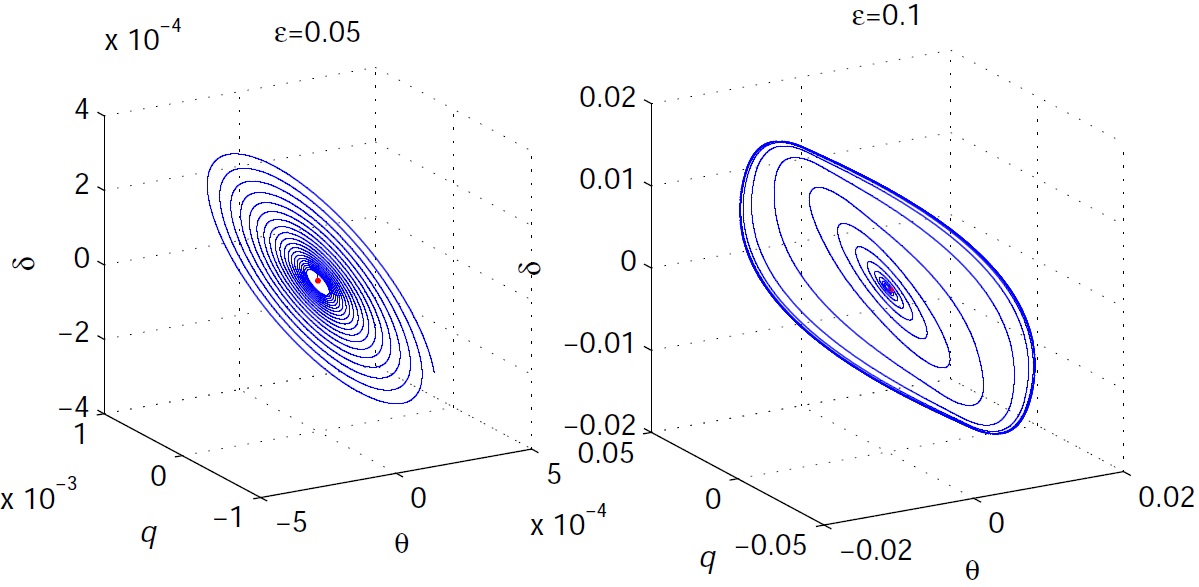}
\includegraphics[width=1\linewidth]{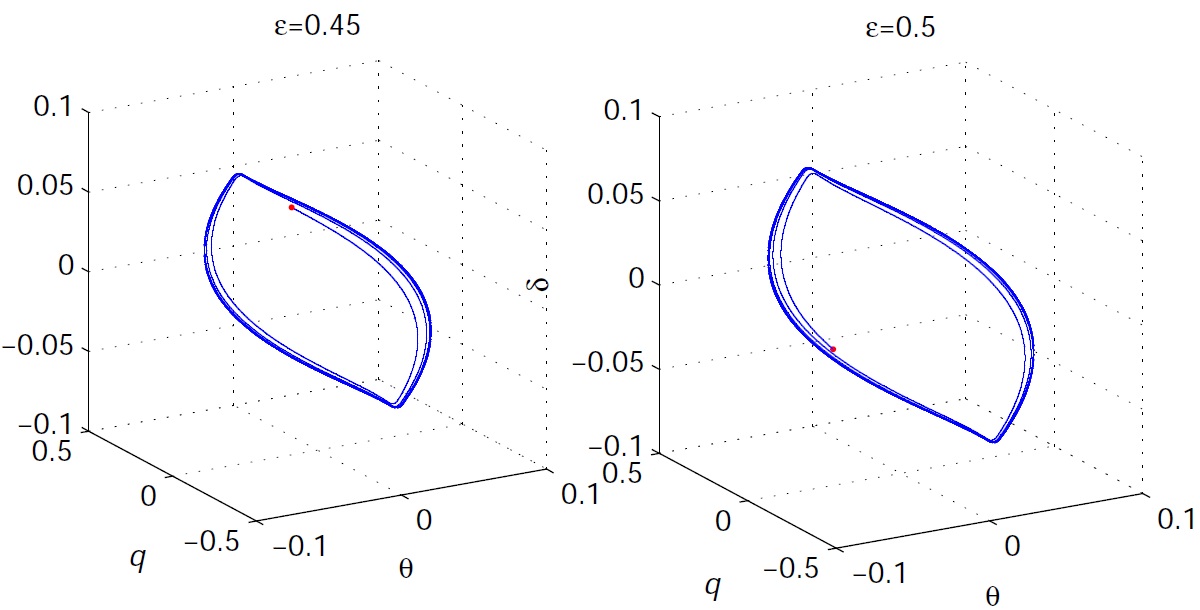}
\includegraphics[width=1\linewidth]{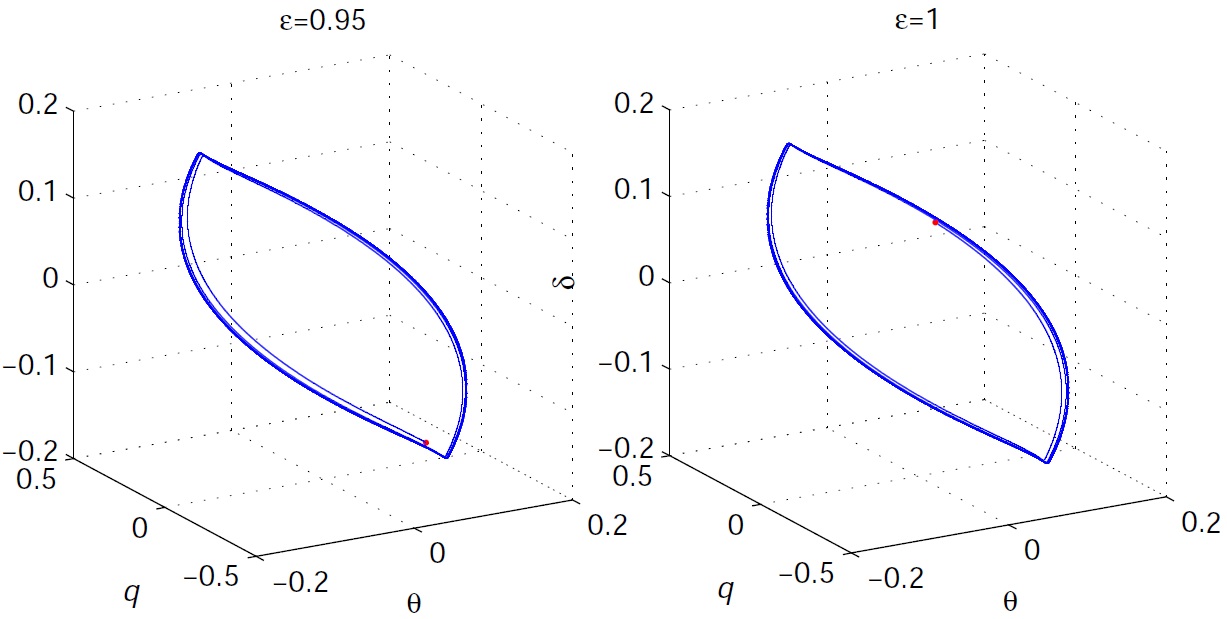}
\caption{Consequent steps of hidden oscillations localization. Phase trajectory projections to subspace $(\theta,q,\delta)$.}
\label{figND3}
\end{figure}

Several consequent steps of hidden oscillations localization via the above procedure
are illustrated by Fig.~\ref{figND3}, where phase trajectory projections to subspace $(\theta,q,\delta)$, where $q$ denotes the pitch rate, for various values of $\eps$
are depicted. Initial value of $x$ is taken as $x_0 =[0,0,0,0,0,0,0,0.1]\trn$.

\begin{figure}[htpb!]
\centering
\includegraphics[width=0.8\linewidth]{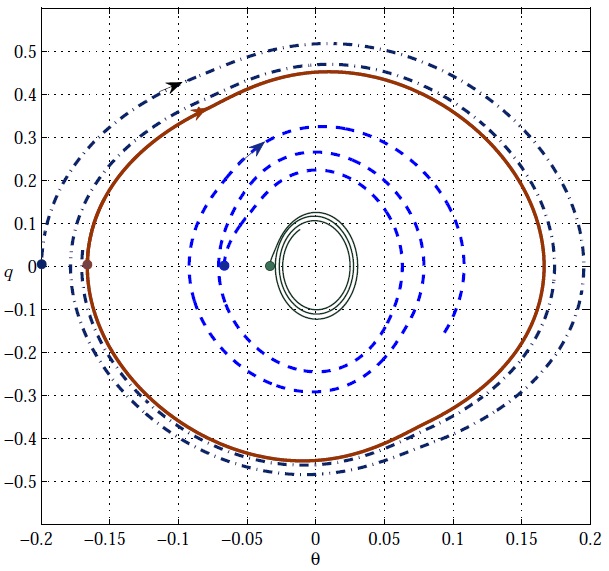}
\caption{Phase trajectory projections to subspace $(\theta,q)$. $x(0) =1.2x_0\trn $ -- dashed line; $x(0) =x_0\trn $ -- thick solid line  (limit cycle);
$x(0) =0.4x_0\trn $ -- thin solid line.}
\label{figND4}
\end{figure}

Finally, the following point $x_0\trn =[0.00264,$$ 0.00292,$$ 4.96\cdot 10^{-1}4,$$
-4.35\cdot 10^{-4}, $$ 7.58\cdot 10^{-5},$$ 6.46\cdot 10^{-5}, 0.0482, 6.45]\trn$  which belongs to the limit cycle has been found after
ten iterations. The simulation results in the form of phase trajectory  projections to subspace $(\theta,q)$ for various initial conditions are plotted in Fig.~\ref{figND4}. Thick solid line on the plot corresponds to the limit cycle oscillation. It is seen that the limit cycle is asymptotically orbitally stable, attracting the neighboring trajectories.
The zero equilibrium is asymptotically stable (thin solid line in Fig.~\ref{figND4}). The simulations demonstrate that the unstable limit cycle for ``intermediate'' initial conditions, separating attractivity to the stable limit cycle oscillations and the equilibrium state may also exist.

\section{Airfoil flutter suppression}\label{Sec:flut}
\subsection{Aeroelastic model of an airfoil section}
An aeroelastic system describes wing dynamics in the presence of a flow field. Interacting forces between the structure, the moment of inertia and air flow destabilize aircraft by producing flutter and limit cycle oscillation \cite{TheodorsenGarrick_TR-685,Theodorsen_NACA35,Bisplinghoff96}. In the presence of a flow field, the wing at a flight speed $U$ oscillates along the plunge displacement direction and
rotates at the pitch angle about the elastic axis. Because flutter
can eventually damage a wing structure, flutter must be shed during a flight.

Let us use the following steady-state model of the airfoil section flutter dynamics \cite{AbdelkefiVasconcellos_ND13,Chen_CNSNS12}:
\begin{align}
&\begin{bmatrix}I_\al&m_wx_\al b\cr m_wx_\al b&m_t \end{bmatrix}
\begin{bmatrix}\ddot \al\cr\ddot h\end{bmatrix} +
\begin{bmatrix}c_\al&0\cr 0&c_h\end{bmatrix}
\begin{bmatrix}\dot \al\cr\dot h\end{bmatrix} \notag\\
&\qquad+
\begin{bmatrix} {k_\al(\al)}&0\cr 0& {k_h(h)}\end{bmatrix}
\begin{bmatrix}\al\cr h\end{bmatrix} =
\begin{bmatrix}M\cr  - L\end{bmatrix}
\label{Chen-1a}\\
& k_\al(\al)=k_1+k_2\al^2,\label{Chen-1}\\
&k_h(h)= \varkappa_1+ \varkappa_2h^2.
\label{Chen-2}
\end{align}
where  $h$ denote the plunge displacement;  $\al$ stands for the pitch angle; $\gamma$ and $\beta$ are angles of the leading and trailing edges, respectively. The wing structure includes a linear
spring oriented along the plunge displacement direction, a rotational spring along the pitch angle, and corresponding dampers.
 $m_t$ denotes is the total weight of the main wing and supporter, $m_w$ is the weight of the main wing, $x_\al$ is the dimensionless distance between the center of mass and the elastic axis, $I_\al$ is the moment of inertia, $b$ is the midchord, $c_\al$ and $c_h$ are the
damping coefficients of the pitch angle and the plunge displacement respectively, $k_h$ and $k_\al(\al)$ are the spring stiffness coefficients
of the plunge displacement and the pitch angle respectively, and $k_\al(\al)$ is a nonlinear term. The nonlinearity $\al k_\al(\al)$ of the
spring, a hard spring in fact, which is actually a hard spring, is defined as

The aerodynamics force  $L$ and torque  $M$ in the low-frequency area and subsonic flight may be represented as follows \cite{Theodorsen_NACA35,TheodorsenGarrick_TR-685,Chen_CNSNS12}:
\begin{align}
&L = \rho U^2bc_{l_\al}s_p\left(\al + \bigg(\dfrac{\dot h}{U} +
\Big(\dfrac{1}{2} - a\Big)\,b\dfrac{\al}{U}\bigg)\right) \notag\\
&\qquad+ \rho U^2bc_{l_\beta}s_p\beta + \rho U^2bc_{l_\gamma}s_p\gamma,
\label{Chen-3}\\
&M = \rho U^2b^2c_{m_{\al\text{-eff}}}s_p\left(\al + \bigg(\dfrac{\dot h}{U} + \Big(\dfrac{1}{2} - a\Big)b\dfrac{\al}{U}\bigg )\right) +\notag\\
&\qquad \rho U^2b^2c_{m_{\beta\text{-eff}}}s_p\beta + \rho U^2b^2c_{m_{\gamma\text{-eff}}}s_p\gamma,
\label{Chen-4}
\end{align}
where $\rho$ is air density, $U$  is the flight speed, $a$ is the dimensionless distance between the elastic axis and the mid-chord; $s_p$ is the windspan length, $c_{l_\al}$, $c_{m_\al}$ are the lift coefficient and moment coefficient per unit angle of attack respectively; $c_{l_\beta}$, $c_{m_\beta}$ are the lift coefficient and moment coefficient per unit angle
respectively against the trailing edge, respectively; $c_{l_\gamma}$, $c_{m_\gamma}$ are the lift coefficient and moment coefficient per unit
angle, respectively, against the leading edge; $c_{m_{\al\text{-eff}}}$, $c_{m_{\beta\text{-eff}}}$, $c_{m_{\gamma\text{-eff}}}$ are the moment derivative coefficient per
unit angle of attack, trailing edge and leading edge, respectively. According to  \cite{Chen_CNSNS12}, they are  defined as
\begin{align}
\begin{array}{l}
c_{m_{\al\text{-eff}}}=\left(\dfrac{1}{2}+a\right)\,c_{l_\al}+2c_{m_\al},\\
c_{m_{\beta\text{-eff}}}=\left(\dfrac{1}{2}+a\right)\,c_{l_\beta}+2c_{m_\beta},\\
c_{m_{\gamma\text{-eff}}}=\left(\dfrac{1}{2}+a\right)\,c_{l_\gamma}+2c_{m_\gamma}.
\end{array}
\label{Chen-5}
\end{align}

Introducing notations $c_1=\rho U^2bs_p$, $c_2=\rho U^2b^2s_p$, rewrite \eqref{Chen-3}, \eqref{Chen-4}  in the form
\begin{align}
\begin{array}{l}
L =c_1\bigg(\al + \bigg(\dfrac{\dot h}{U} +
\Big(\dfrac{1}{2} - a\Big)\,b\dfrac{\al}{U}\bigg)\bigg)\\
\qquad+c_1c_{l_\beta} {\beta} + c_1c_{l_\gamma} {\gamma},
\\
M=c_2c_{m_{\al\text{-eff}}}\bigg(\al + \bigg(\dfrac{\dot h}{U} + \Big(\dfrac{1}{2} - a\Big)b\dfrac{\al}{U}\bigg )\bigg)+\\
\qquad c_2c_{m_{\beta\text{-eff}}} {\beta}+c_2c_{m_{\gamma\text{-eff}}} {\gamma}.
\end{array}
\label{Chen-6}
\end{align}
\begin{figure}
\centering
\includegraphics[width=\linewidth]{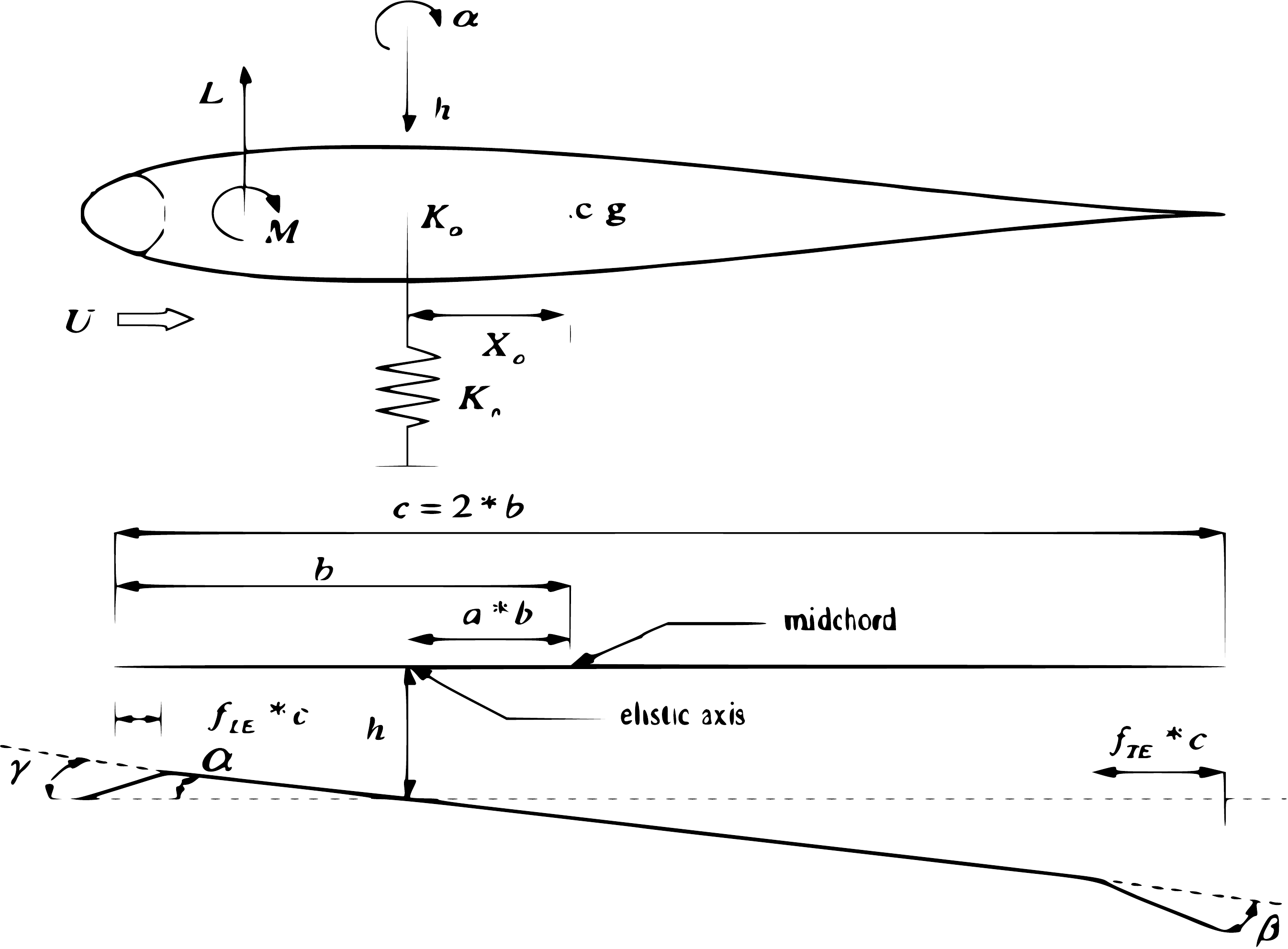}
\caption{The airfoil section with controlling surfaces (cf. \cite{Chen_CNSNS12}).}
\label{Chen_CNSNSf1}
\end{figure}
Denote the state-space vector as ${\bf x}=\begin{bmatrix}\al& \dot\al&h&\dot h\end{bmatrix}\trn$; $k_h(h)\equiv k_h$. Then
\begin{align}
\begin{cases}
\dot x_1=x_2,\\
\dot x_2=c_{\al_1}x_1+c_{\al_{\text{nonl}1}} {x_1^3}+c_{\dot{\al}_1}x_2
+c_{h_1}x_3+c_{\dot{h}_1}x_4\\\qquad+c_{\beta_1} {\beta}+c_{\gamma_1} {\gamma},\\
\dot x_3=x_4,\\
\dot x_4=c_{\al_2}x_1+c_{\al_{\text{nonl}2}} {x_1^3}+c_{\dot{\al}_2}x_2
+c_{h_2}x_3+c_{\dot{h}_2}x_4\\\qquad+c_{\beta_2} {\beta}+c_{\gamma_2} {\gamma},
\end{cases}
\label{Chen-6a}
\end{align}
where $c_{\al_1}$, $c_{\al_{\text{nonl}1}}$, $c_{\dot{\al}_1}$, $c_{h_1}$, $c_{\dot{h}_1}$, $c_{\beta_1}$, $c_{\gamma_1}$, $c_{\al_2}$, $c_{\al_{\text{nonl}2}}$, $c_{\dot{\al}_2}$, $c_{h_2}$, $c_{\dot{h}_2}$, $c_{\beta_2}$, $c_{\gamma_2}$ are model parameters which are assumed to be constant on the considered time interval.

\begin{figure}[htpb!]
\centering
\includegraphics[width=0.8\linewidth]{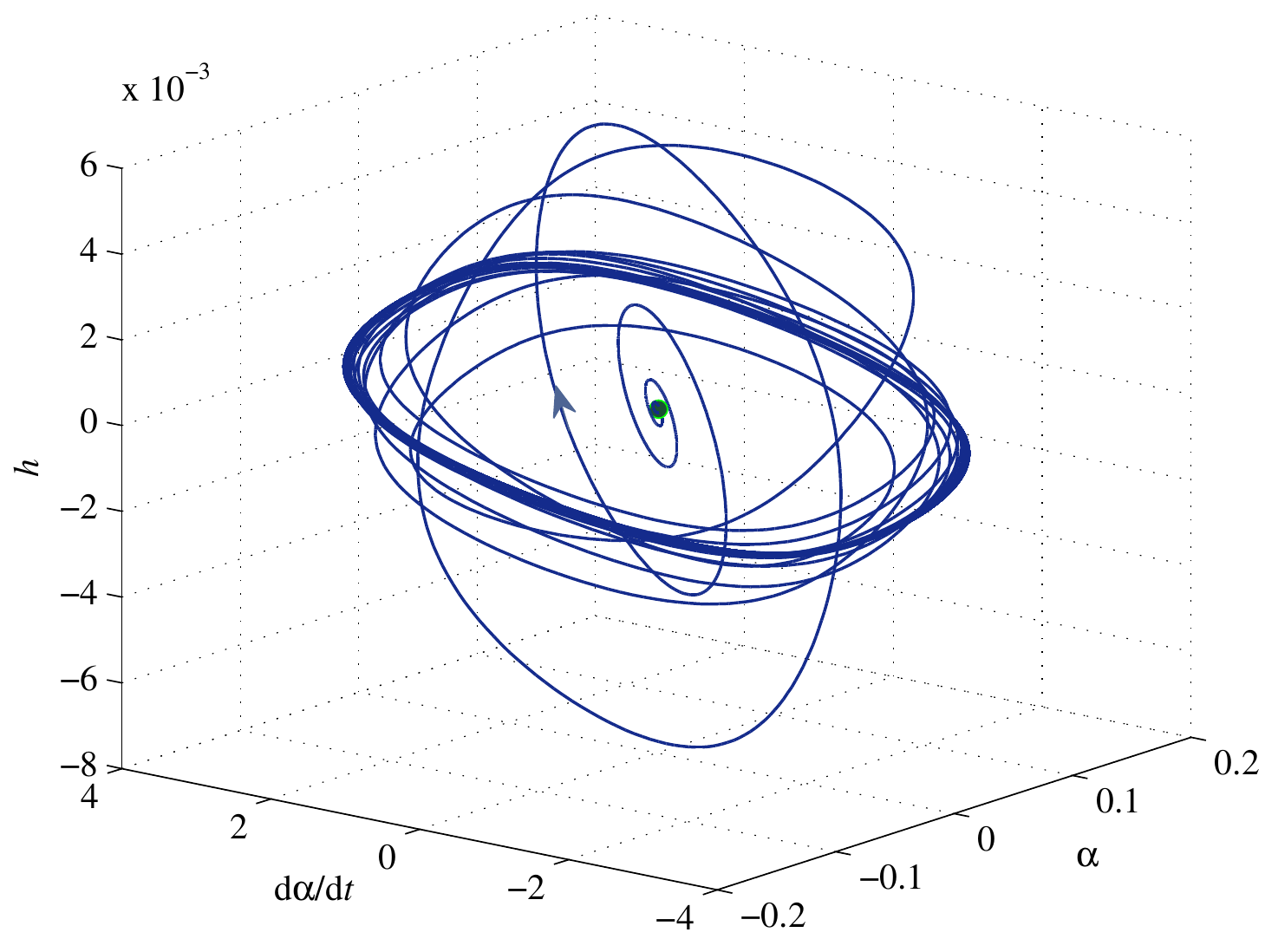}
\caption{Free motion. Limit cycle oscillation birth; projection of phase trajectory to subspace $(\al,\dot \al, h)$. Initial point $\al(0)=0.1$~deg, $\dot\al(0)=h(0)=\dot h(0)=0$.}
\label{lco_free}
\end{figure}

Let us linearize \eqref{Chen-6a} in the vicinity of the origin and represent it in the vector-matrix form as $\dot {\bf x}(t)={\bf A}{\bf x}(t)+{\bf b}u(t)$, where ${\bf A}$ is $(4\times 4)$ system matrix,  ${\bf b}$ is $(4\times 1)$ input matrix, $u(t)\equiv \beta(t)$ denotes the control action (the controlling surface deflection).  Let the airfoil model \eqref{Chen-6a} parameters be taken as in \cite{Chen_CNSNS12}, see Appendix~\ref{appaer}. Matrix ${\bf A}$ eigenvalues are as $ s=\{3.05\pm 15i,-4.63\pm 13.5i\}$ which shows that system \eqref{Chen-6a}  origin is unstable in the Lyapunov sense. Meanwhile, as shown in the series of papers, cf. \cite{PriceAlighanbari_JFS95,ONeilStrganac_JoA98,Ding_AST06,LeeLeblanc86,ZhangWen_AMC13,PriceLee_JA95,AlighanbariPrice_ND96,KoStrganacKurdila_ND99,Highly_Accurate_Cycle,ADA384971,KoStrganacJunkins_JVC02,LiGuo_JSV10,Chen_CNSNS12,AbdelkefiVasconcellos_ND13,BichiouHajjNayfeh_ND16,IraniSazesh_ND16,HeYang_ND17,TianYangGu_ND17,WeiMottershead_ASTE17,FazelzadehAzadi_JVC17},  due to presence of the cubic nonlinearity in the system model, the system trajectories are bounded and the limit cycle oscillations (the airfoil flutter phenomenon) arise. An illustration of the limit cycle oscillations birth for small initial conditions is given by Fig.~\ref{lco_free}, where projection of the free motion phase trajectory to subspace $(\al,\dot \al, h)$ for $\al(0)=0.1$~deg, $\dot\al(0)=h(0)=\dot h(0)=0$ is plotted.

\subsection{Airfoil flutter suppression system}

Consider the problem of flutter suppression by controlling the airfoil trailing edge  $\beta$  (i.e. assume   hereafter that $\gamma\equiv 0$).

There is a plenty of papers  devoted to airfoil flutter active suppression systems, see \cite{edwards1977unsteady,StrganacKo_AIAA99,KoStrganacKurdila_ND99,KoStrganacJunkins_JVC02,LiGuo_JSV10,Chen_CNSNS12,Piovanelli_ECC16,BichiouHajjNayfeh_ND16,WeiMottershead_ASTE17,FazelzadehAzadi_JVC17} for mentioning a few. To simplify the exposition by avoiding unnecessary difficulties in control law synthesis, in the present study the widespread LQR-control design technique is employed.

Consider the static state feedback control law $u=-K_\text{fb}{\bf x}\inr^4$, where state-space vector ${\bf x}$, as in \eqref{Chen-6a}, is a measured plant state, $K_\text{fb}$ is $(1\times 4)$ matrix (row vector) of the controller parameters.  Introduce the linear-quadratic performance index  $J = \int_0^\infty{\big(x\trn Qx + u\trn Ru}\big)\,{\rm d}t$ with a given positively defined  $(4\times 4)$ matrix $Q=Q\trn>0$ and a non-negative scalar $R\ge 0$. Employing the standard MATLAB routine {\sl lqr} one obtains the state feedback vector $K_\text{fb}$, minimizing performance  index $Q$.

\begin{figure}[htpb!]
\centering
\includegraphics[width=0.8\linewidth]{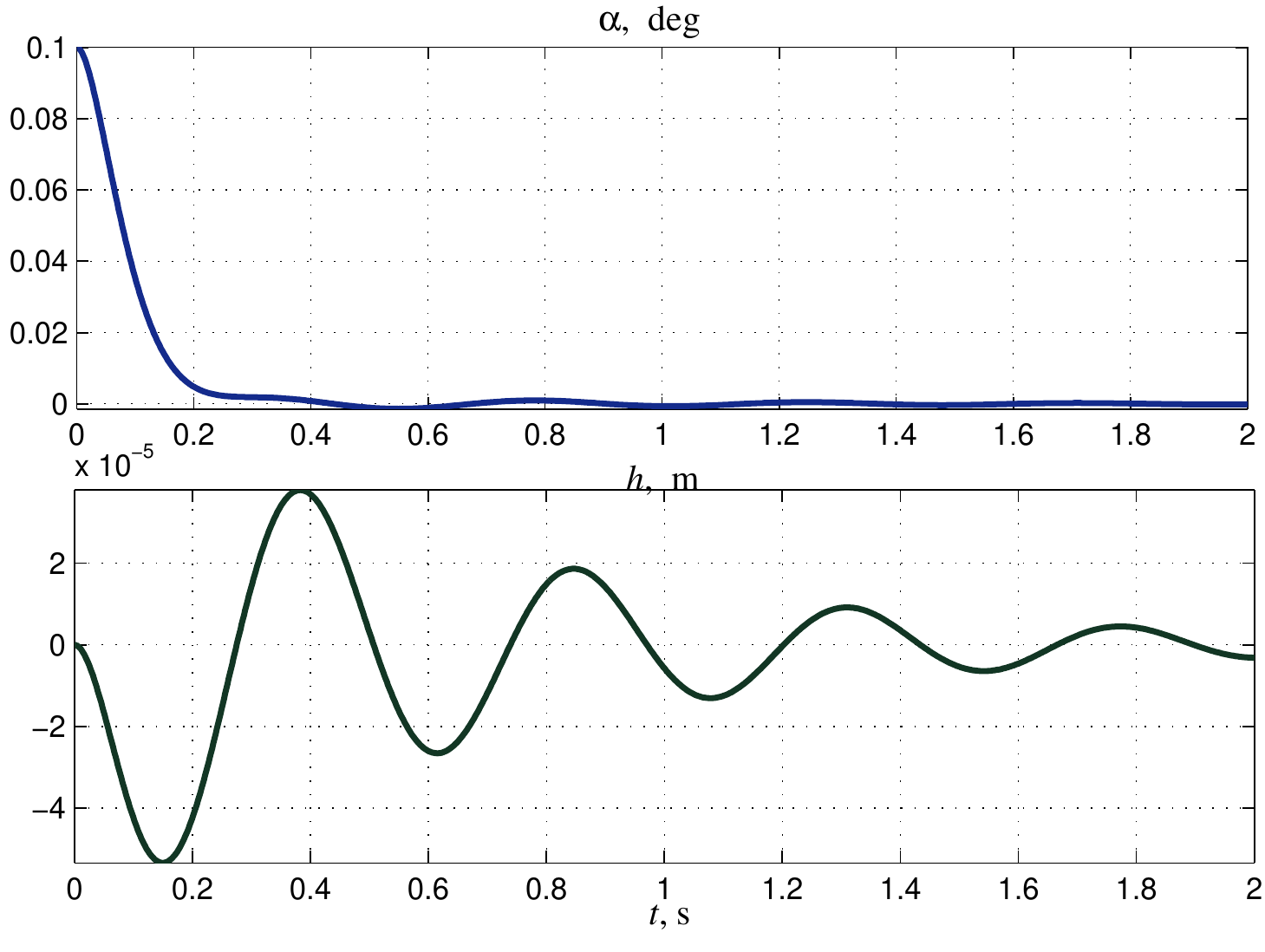}
\caption{Time histories of $\al(t)$, $h(t)$ for controlled motion. ``Ideal'' static state feedback controller. Initial point $\al(0)=0.1$~deg, $\dot\al(0)=h(0)=\dot h(0)=0$.}
\label{transcontrid}
\end{figure}

Let us pick up $Q=\diag\{1,0.01,1,2\cdot 10^{-3}\}$, $R=0.5$.  MATLAB  linear-quadratic optimization  routine {\sl lqr} leads then to the following state feedback vector $K_\text{fb}=\begin{bmatrix}-0.93& -0.17& -7.22&  0.062\end{bmatrix}]$. This gives the closed-loop linearized system matrix ${\bf A}_\text{fb}={\bf A}-{\bf b}K_\text{bf}$ with the eigenvalues $ s_\text{fb}=\{ -17.6\pm 9.0i, -1.53 \pm 13.6i\}$, ensuring asymptotic convergent system behavior in the close vicinity of the origin. Time histories of $\al(t)$, $h(t)$ for controlled motion for the case of the ``ideal'' static state feedback controller at the initial point $\al(0)=0.1$~deg, $\dot\al(0)=h(0)=\dot h(0)=0$ are depicted in Fig.~\ref{transcontrid}. The case of dynamical actuator with the magnitude and rate limitations is considered below.

Since we focus our attention to existence of hidden oscillation rather than to controller design problems, for simplicity, in the above control law synthesis, the
actuator dynamics have been omitted and the assumption that $\beta(t)\equiv u(t)$ was adopted. Let us
 check the control system performance taking into account the actuator
model. To this end a second-order actuator model with the output magnitude and rate saturations \eqref{bia32b}, described in Sec.~\ref{Sec:modact} is used where $\delta$, $\dot \delta$,  $\bar \delta$ $\bar{\dot \delta}$ are substituted by $\beta$, $\dot \beta$,  $\bar \beta$ $\bar{\dot \beta}$ respectively.
In the present study is taken that  $\bar \beta=0.0873~\text{rad}=5$~deg, $\bar{\dot \beta}=8.73~\text{rad/s}=500$~grad/s. $\xi_\text{act}=0.6$, $\omega_\text{act}=50$~rad/s. Gain $\lambda$ in \eqref{bia32b} is set to $100$.

\begin{figure}[ht]
\centering
\includegraphics[width=0.9\linewidth]{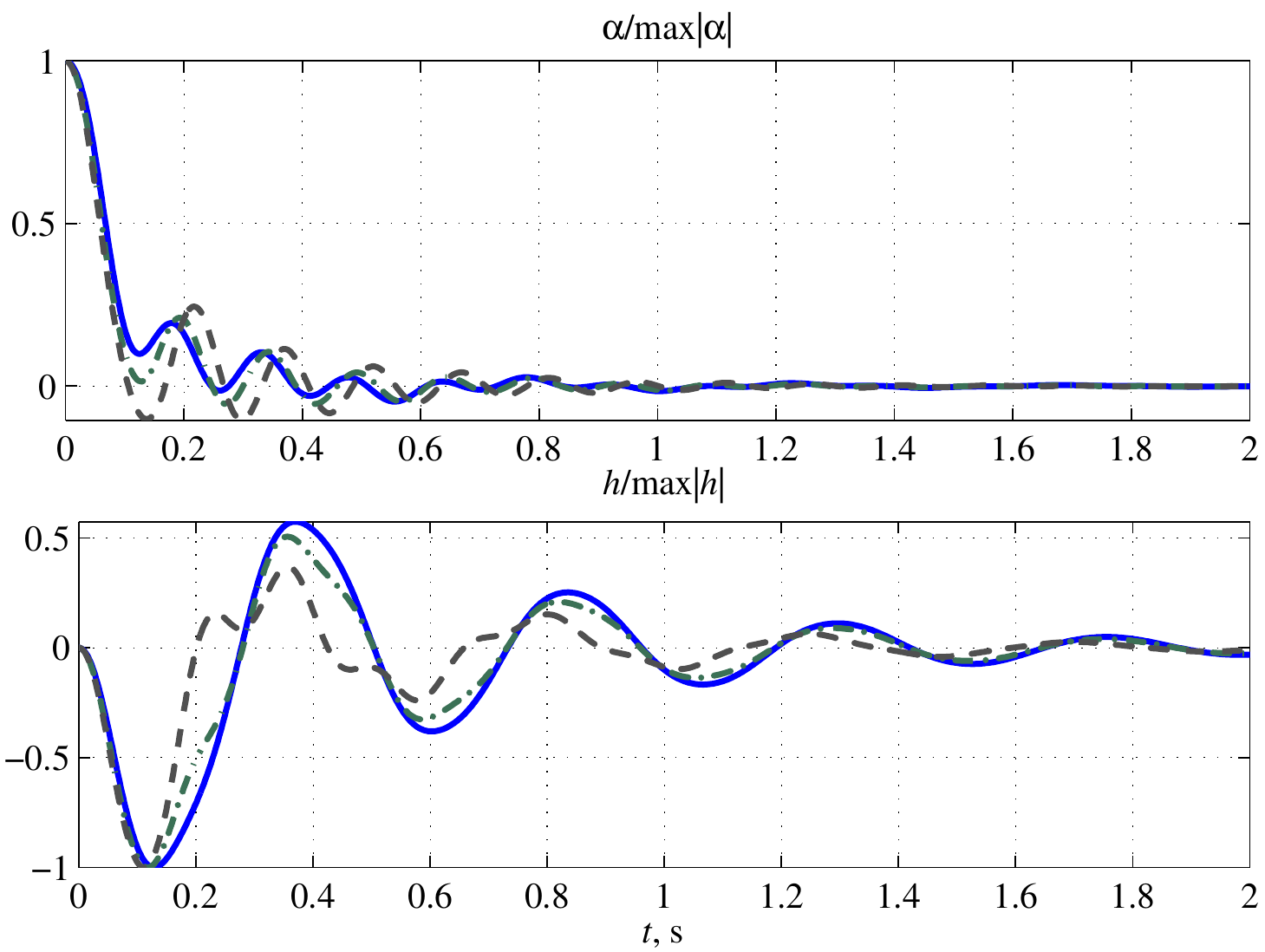}
\caption{Time histories of $\al(t)/\max_t|\al|$ (upper plot), $h(t)/\max_t|h|$ (lower plot). $\al(0)=0.1$~deg -- solid line, $\al(0)=4$ deg -- dash-dot line, $\al(0)=5$ deg -- dashed line.}
\label{3timihist}
\end{figure}

\begin{figure}[htp]
\centering
\includegraphics[width=0.8\linewidth]{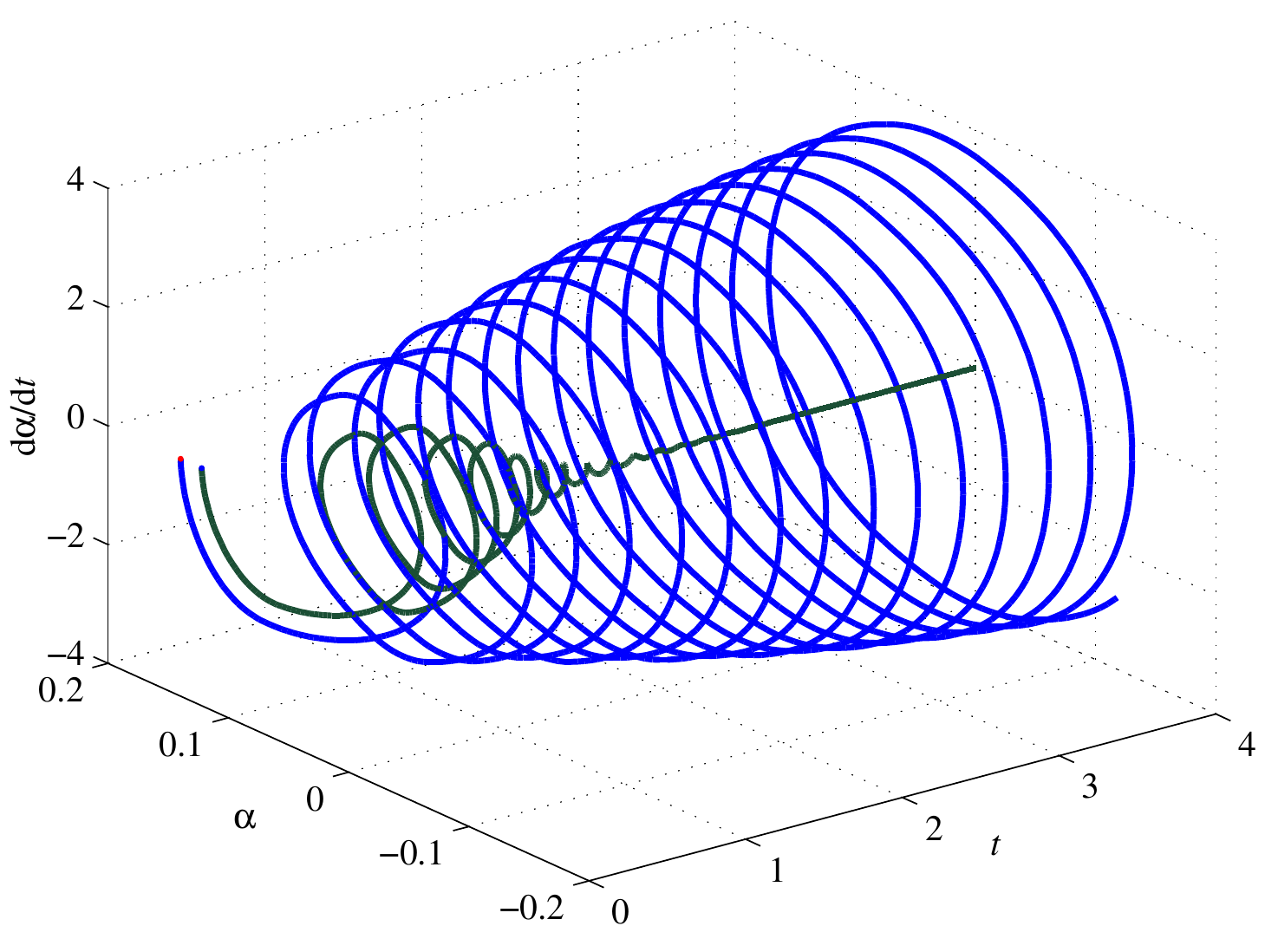}
\caption{System motion in the subspace $(t,\al,\dot\al)$ for $\al(0)=7$~deg (trajectories converge to zero) and $\al(0)=8$~deg (limit cycle oscillation arises). }
\label{chenalintcur}
\end{figure}

\begin{figure}[htp]
\centering
\includegraphics[width=0.9\linewidth]{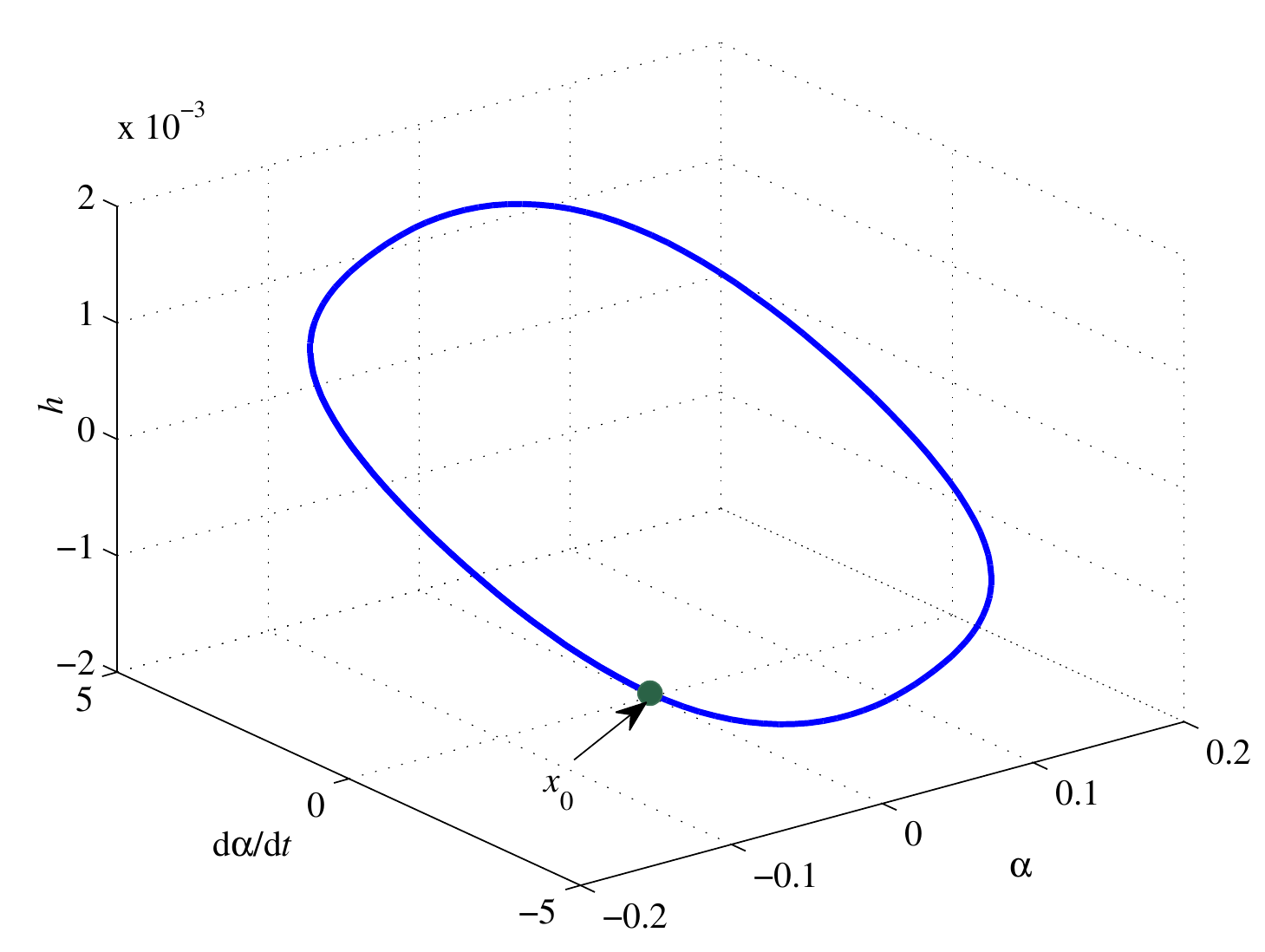}
\caption{Limit cycle oscillation in the subspace $(\al,\dot \al, h)$. Initial point  $\al(0)=-0.109$ rad, $\dot\al(0)= -3.55$ rad/s, $h(0)=-9.33\cdot 10^{-4}$ m, $\dot h(0)=0.031$ m/s, $\beta(0)= -0.0873$ rad, $\dot\beta(0)= -8.723$ rad/s.}
\label{lco_aldalh}
\end{figure}

\balance
\subsection{Localization of hidden oscillation}
The numerical evaluation shows that in the linear approximation, the actuator dynamics do not effect significantly the closed-loop system performance, as is seen from the plots of Fig.~\ref{3timihist}. In this figure, time histories of variables $\al(t)$, $h(t)$ for various $\al(0)$  (the other initial state variables are set to zero), normalized with respect to their maximal values  are plotted. Comparing the curves depicted on Fig.~\ref{transcontrid} with the corresponding (solid) curves of Fig.~\ref{3timihist}, one may notice that in the linear consideration of actuator dynamics (when the  saturations are not active), the unmodelled actuator dynamics  \eqref{bia31} do not lead to significant degradation of the overall system performance. The case of $\al(0)\approx 4$~deg may be treated as a ``boundary'' one, when the nonlinearities become active. If the initial conditions are sufficiently large, the feedback controller looses its stabilizing properties and the limit cycle oscillation appears. This oscillation can not be found by linearization of the system model in the vicinity of the  origin and, therefore, may be referred to as a ``hidden'' one. This phenomenon is illustrated by Fig.~\ref{chenalintcur}, where the system trajectories in the subspace $(t,\al,\dot\al)$ for $\al(0)=7$~deg and $\al(0)=8$~deg are plotted. In the first case they tend to the origin, in the second one the limit circle oscillation occurs. By applying the hidden oscillations localization procedure of Sec. \ref{Sec:analmeth}, the initial point of limit cycle oscillation is found as  $\al(0)=-0.109$ rad, $\dot\al(0)= -3.55$ rad/s, $h(0)=-9.33\cdot 10^{-4}$ m, $\dot h(0)=0.031$ m/s, $\beta(0)= -0.0873$ rad, $\dot\beta(0)= -8.723$ rad/s. The projection of the limit cycle phase plot to the subspace $(\al,\dot \al, h)$ is plotted in Fig.~\ref{lco_aldalh}.

\section{Conclusions}\label{Sec:conc}
In the paper, the control problem with limitations on the magnitude and rate of the control action in  aircraft control systems, is studied.
Existence of hidden limit cycle oscillations in the
case of actuator position and rate limitations  is demonstrated by the examples of piloted aircraft
PIO phenomenon and the airfoil flutter suppression system. Hidden oscillations in the pilot-aircraft loop are studied and localized by means of the iterative analytical-numerical method.

\section*{Acknowledgments}
This work was supported by Russian Science Foundation project (14-21-00041).

\section*{Appendix. Aeroelastic model parameters}\label{appaer}

\begin{table}[ht]
\caption{Values of initial parameters of aeroelastic model \eqref{Chen-4}, \eqref{Chen-5}.}
\label{Chenparmod}
\begin{tabular}{ll|ll}
\hline\noalign{\smallskip}
$a$&$-0.6719$&$c_{m_\gamma}$&$-0.1005$\\
$b$&$0.1905$~m&$I_\al$&$(m_wx_\al^2 b^2+0.009039)$~kg m$^2$\\
$c_\al$&$0.036$~kg m$^2$/s&$k_\al(\al)$&$12.77+1003\al^2$\\
$c_h$&$27.43$~kg/s&$k_h$&$2844.4$~N/m\\
$c_{l_\al}$&$6.757$&$m_t$&$15.57$~kg\\
$c_{l_\beta}$&$3.358$&$m_w$&$4.34$~kg\\
$c_{l_\gamma}$&$-0.1566$&$s_p$&$0.5945$~m\\
$c_{m_\al}$&$0$&$x_\al$&$-(0.0998+a)$\\
$c_{m_\beta}$&$-0.6719$&$\rho$&$1.225$~kg/m$^3$\\
\noalign{\smallskip}\hline
\end{tabular}
\end{table}

\begin{table}[ht]
\caption{Simulation model parameters.}
\label{Chenval}
\begin{tabular}{ll|ll|ll}
\hline\noalign{\smallskip}
$U$&$19.0625$&$m_w$&$4.3400$&$a$&$-0.6719$\\
$x_\al$&$0.5721$&$c_{m_\gamma}$&$-0.1005$&$b$&$0.1905$\\
$I_\al$&$0.0606$&$c_\al$&$0.0360$&$k_1$&$12.77$\\
$k_2$&$1003$&$c_h$&$27.4300$&$k_h$&$2.844\cdot 10^3$\\
$c_{l_\al}$&$6.7570$&$m_t$&$15.57$&$c_{l_\beta}$&$3.3580$\\
$c_{l_\gamma}$&$-0.1566$&$s_p $&$0.5945$&$c_{m_\al}$&$0$\\
$c_{m_\beta}$&$-0.6719$&$\rho$&$1.2250$&$c_{m_{\al\text{-eff}}}$&$-1.1615$\\
$c_{m_{\beta\text{-eff}}}$&$-1.9210$&$c_{m_{\gamma\text{-eff}}}$&$-0.1741$&$c_1$&$50.4130$\\
$c_2$&$ 9.6037$&$c_{\al_1} $&$-211.39$&$c_{\al_{\text{nonl}_1}}$&$-778.5$\\
$c_{\dot{\al}_1}$&$ -0.7076$&$c_{h_1} $&$1.3454\cdot 10^3$&$c_{\dot {h}_1}$&$12.3153$\\
$c_{\beta_1} $&$-207.1799$&$c_{\gamma_1}$&$-29.7643$&$c_{\al_2} $&$-9.32 25$\\
$c_{\al_{\text{nonl}_2}} $&$  23.6498$&$c_{\dot {\al}_2} $&$   -0.1629$&$c_{h_2} $&$ -172.3376$\\
$c_{{\dot h}_2} $&$ -2.4678$&$c_{\beta_2} $&$-1.5305$&$c_{\gamma_2} $&$ 1.2691$\\
\noalign{\smallskip}\hline
\end{tabular}
\end{table}

Model \eqref{Chen-6a} parameters are defined by the following expressions, see \cite{Chen_CNSNS12}:
\begin{align*}
\begin{array}{l}
c_{\al_1}=c_2m_tc_{m_{\al\text{-eff}}}+c_1m_wx_\al bc_{l\al}-m_tk_1,\\
c_{\al_{\text{nonl}1}}=-m_tk_2,
c_{\dot{\al}_1}=c_2m_tc_{m_{\al\text{-eff}}}\left(\dfrac{1}{2}-a\right)b\dfrac{1}{U}\\\quad+c_1m_wx_\al bc_{l_\al}\left(\dfrac{1}{2}-a\right)b\dfrac{1}{U}-c_\al m_t,
c_{h_1}=k_hm_wx_\al b,\\
c_{\dot{h}_1}=c_2m_tc_{m_{\al\text{-eff}}}\dfrac{1}{U}+c_1m_wx_\al bc_{l_\al}
\dfrac{1}{U}+c_hm_wx_\al b,\\
c_{\beta_1}=c_2m_tc_{m_{\beta\text{-eff}}}+c_1m_wx_\al bc_{l_\beta},\\
c_{\gamma_1}=c_2m_tc_{m_{\gamma\text{-eff}}}+c_1m_wx_\al bc_{l_\gamma},\\
c_{\al_2}=-c_2m_wx_\al bc_{m_{\al\text{-eff}}}-c_1I_\al c_{l_\al}+m_wx_\al bk_1,\\
c_{\al_{\text{nonl}2}}=m_wx_\al bk_2,\\
c_{\dot{\al}_2}=-c_2m_wx_\al bc_{m_{\al\text{-eff}}}\left(\dfrac{1}{2}-a\right)b\dfrac{1}{U}\\
\quad-c_1I_\al c_{l_\al}\left(\dfrac{1}{2}-a\right)b\dfrac{1}{U}+c_\al m_wx_\al b,\\
c_{h_2}=-k_hI_\al,
\end{array}
\end{align*}

\begin{align}
\begin{array}{l}
c_{\dot{h}_2}=-c_2m_wx_\al bc_{m_{\al\text{-eff}}}\dfrac{1}{U}
-c_1I_\al c_{l_\al}\dfrac{1}{U}-c_hI_\al,\\
c_{\beta_2}=-c_2m_wx_\al bc_{m_{\beta\text{-eff}}}-c_1I_\al c_{l_\beta},\\
c_{\gamma_2}=-c_2m_wx_\al bc_{m_{\gamma\text{-eff}}}-c_1I_\al c_{l_\gamma}.
\end{array}
\label{Chen_par-2}
\end{align}

The  aeroelastic model \eqref{Chen-4}, \eqref{Chen-5} parameters are taken as in  \cite{Chen_CNSNS12}, see Tab.~\ref{Chenparmod}. Calculations according to \eqref{Chen-5},  \eqref{Chen_par-2} lead to model \eqref{Chen-6a} parameter values, given in Tab.~\ref{Chenval}.

\end{document}